# Polarization-Engineered InGaN/GaN Heterojunctions for Photovoltaic Applications

S. A. Kazazis [1,*], E. Papadomanolaki [1], and E. Iliopoulos [1,2]

[1] *Department of Physics, University of Crete, P.O. Box 2208, 71003 Heraklion, Greece*

[2] *Microelectronics Research Group, IESL-FORTH, P.O. Box 1385, 71110 Heraklion, Greece*

**Abstract**

The photovoltaic properties of (0001) *n*-InGaN/*p*-GaN single heterojunctions were investigated numerically and compared with those of conventional *p*-GaN/i-InGaN/*n*-GaN structures, employing realistic material parameters. This alternative device architecture exploits the large polarization fields, and high efficiency modules are achieved for In-rich, partially relaxed and coherently strained InGaN films. Conversion efficiencies up to 14% under AM1.5G illumination can be reached, revealing the true potential of InGaN single junction solar cells with proper design.

\* Corresponding author: kazazis@physics.uoc.gr

## 1. INTRODUCTION

Ternary indium-gallium-nitride (InGaN) alloys are the main building blocks for light-emitting diodes (LEDs) and laser diodes. Their inherent properties, such as their direct band gap, tunable from 0.64 to 3.38 eV [1], [2], high absorption coefficient [3], [4] and radiation resistance [5] also make InGaN compound semiconductors an excellent candidate for photovoltaic applications. Despite the superior properties of InGaN, there are still many roadblocks to achieving high-efficiency solar cells. One bottleneck limiting the performance of such devices arises from the potential barrier in the GaN/InGaN hetero-interface due to the electron affinity difference between InN and GaN [6]. Another important factor influencing the photovoltaic properties of III-nitride solar cells is the existence of significant interface charges induced by spontaneous and piezoelectric polarizations [7]–[9]. A method to overcome these constraints is by using InGaN homojunction structures instead of InGaN/GaN heterostuctures. Although InGaN homojunction devices are predicted to achieve greater photovoltaic properties than their heterojunction counterparts [10], experimental evidence proved the opposite mainly due to the difficulty of growing high quality *p*-InGaN layers [11]. It was not until very recently that *p*-i-*n* InGaN homo-junctions with high indium content and reduced stacking fault density were reported [12], underling that there is still a lot to be done towards efficient InGaN-only photovoltaics.

Until now, most of the research development on InGaN solar cells is based on the fundamental double-heterojunction design of III-nitride LEDs, consisting of a bulk intrinsic InGaN layer or GaN/InGaN multiple quantum wells grown on *n*-type GaN, capped with a *p*-GaN layer [13]–[18]. In such configurations, the polarization charges at the hetero-interfaces generate a huge electrostatic field, whose direction is against the built-in electric field of the junction and, consequently, the efficient collection of photo-generated carriers is impeded [7], [8], [19]. Thus, the most challenging issue dealing with *p*-on-*n* structures is to mitigate the polarization-induced electric field. The polarization effect can be effectively suppressed by inserting step-graded [20] or compositional graded layers between hetero-interfaces [21], but both approaches require highly doped ($5\times10^{18}$ cm$^{-3}$) *p*-type InGaN regions, which is practically very difficult to realize, especially in the case of compositional graded *p*-InGaN.

As an alternative to avoid the abovementioned limitations and exploit the polarization effect, the use of *n*-i-*p* structure instead of the conventional *p*-i-*n* structure has been proposed [9], [22]. One major advantage of the *p*-side down architecture is that the polarization-induced electric field is in favor of the efficient carrier extraction, not against it, leading to high device performance. Such architectures using the polarization effect have already been employed at Ga-polar III-nitride LEDs [23], but on InGaN based solar cells only a few reports exist, focusing only on specific aspects of this design from a theoretical point of view. In this paper, to shed light on the non-conventional *p*-side down design and explore its potential for high-efficiency InGaN-based photovoltaic devices, Ga-face *n*-InGaN/*p*-GaN heterojunctions were investigated numerically and compared with the conventional *p*-GaN/i-InGaN/*n*-GaN structures, using realistic material parameters, attainable with the current status of InGaN alloy epitaxy.

## 2. SIMULATION PARAMETERS

In the simulations, Poisson equation, current continuity equations, scalar wave equation and photon rate equation, were solved self-consistently, employing the finite-element analysis software APSYS [24]. For the calculation of the energy bands the 6×6 k · p model developed by Chuang and Chang for strained wurtzite semiconductors [25], [26] was considered, whereas incident light transmission and absorption were treated using transfer matrix method. Bound sheet charge density, induced by spontaneous and piezoelectric polarization, for single and double heterojunctions was calculated according to Fiorentini *et al.* [27] depending on the strain state of the films, while, in the case of graded composition layer the immobile volume charge density was extracted from the differential form of Gauss's law [28]. The strain state in the simulations, and consequently the amount of fixed polarization charges, was described by the quantity R which represents the percentage of plastic relaxation of the films. The values of relaxation (R) studied were 100% for the fully relaxed case, 50% for partially relaxed and 0% for coherently strained layers.

The values of the dielectric constant and electron (hole) effective masses for GaN and InN are taken from [6] and their corresponding values for InGaN were determined using a linear interpolation between GaN and InN values. Regarding minority carrier lifetime, experimental values reported for high quality bulk GaN and InN are in the range of 6 and 5 ns, respectively [29], [30]. Due to the lack of experimental data for the minority carrier lifetime of InGaN alloys, the realistic and widely used [6], [7], [9], [10], [21], [22] in theoretical studies in InGaN-based solar cells value of 1 ns was assumed. In the simulations, the effect of radiative and Auger recombination is also considered. The radiative recombination coefficient values are calculated by linear interpolation between GaN and InN values taken from [31] and [32], respectively, whereas the Auger recombination rates for electrons and holes are taken from [33]. The doping dependent electron and hole mobilities for GaN and InN were calculated using the well-known Caughey-Thomas approximation [34]

$$\mu_i(N) = \mu_{\min,i} + \frac{\mu_{\max,i} - \mu_{\min,i}}{1 + \left(\frac{N}{N_{g,i}}\right)^{a_i}} \tag{1}$$

where $i = n, p$ for electrons or holes respectively, and $N$ is the carrier concentration. The parameters $\mu_{max}$, $\mu_{min}$, $N_{g,i}$ and $\alpha_i$ depend on the type of the semiconductor material and their values for GaN and InN can be found in [6]. For the InGaN case, mobilities were also determined by linear interpolation between GaN and InN values.

Among the parameters influencing the performance of a solar cell, band gap energy, absorption coefficient and refractive index are of outmost importance. In order to simulate InGaN-based photovoltaic devices more realistically, $In_xGa_{1-x}N$ alloys in the entire composition range were epitaxially grown on commercial free-standing GaN by plasma-assisted molecular beam epitaxy. The growth conditions of the films can be found in [35]. The InN mole fraction (MF) was determined by high resolution X-ray diffraction measurements using a Bede D1 triple-axis X-ray diffractometer, whereas the optical dielectric functions (absorption spectra and refractive index dispersion relations) of the alloys were obtained from the analysis of the spectroscopic ellipsometry data acquired using a rotating analyzer J.A. Woolam VASE. The dielectric functions of the InGaN thin films were modeled employing Herzinger-Johs critical point generalized oscillator functions in order to achieve a Kramers–Kronig consistent description. Further information about the ellipsometric data analysis is beyond the scope of the present paper and will be reported elsewhere. The resulting absorption coefficient spectra as well as the wavelength-dependent refractive indices of GaN and $In_xGa_{1-x}N$ films used in the simulations are presented in Fig. 1.

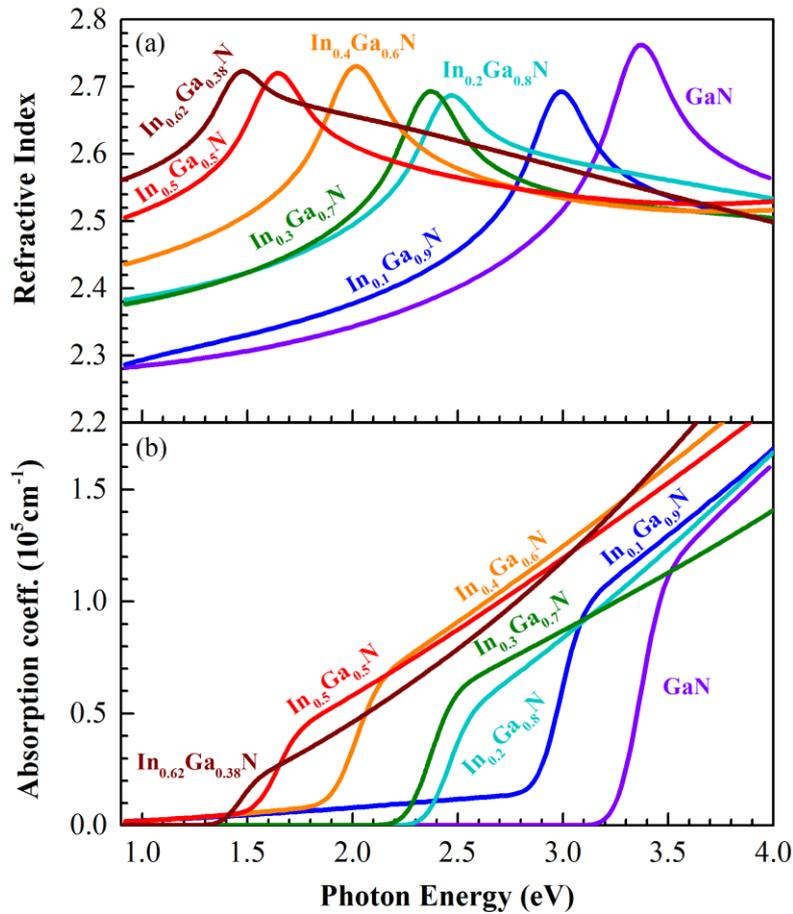

Fig.1. (a) Refractive index dispersion relations and (b) absorption coefficient spectra of GaN substrate and InxGa1-xN films used in the simulations, as extracted from the ellispometric data analysis.

Also, from the abovementioned analysis the bandgap energy of the films was derived and it was found that the composition dependence of the strain-free band gap at room temperature in the entire composition range is well expressed by a bowing parameter of $b=1.67\pm0.09$ eV. This value is in good agreement with the ab initio calculated bandgap dependence for uniform (not clustered) InGaN alloys [36]. Finally, the conduction/valence band-offset ratio was set to be 0.7/0.3 [37].

## 3. RESULTS AND DISCUSSION

### A. Comparison Between p-i-n and n-p Structures

The typical $p$-GaN/i-InGaN/$n$-GaN solar cell structure investigated in the simulations was slightly different than the one reported in [16], since the doping concentration of the bottom $n$-GaN in the cases studied was $1\times10^{18}$ cm$^{-3}$ instead of $2\times10^{18}$ cm$^{-3}$. The carrier concentration of the 200 nm i-InGaN was chosen to be $5\times10^{16}$ cm$^{-3}$ and $5\times10^{17}$ cm$^{-3}$, since it is difficult to grow InGaN layers with low background electron concentration due to its strong propensity to be unintentionally $n$-type doped [38], while the InN MF spanned a range of 0.1 to 0.4. Regarding the $n$-$p$ structure, it was consisted of a 2 $\mu$m thick $p$-type GaN substrate followed by a 200 nm $n$-InGaN layer on the top. The hole concentration of the $p$-GaN was chosen to be $5\times10^{17}$ cm$^{-3}$, which is a typical value for commercial $p$-type GaN substrates. InGaN doping levels and InN MF investigation range remained the same as the conventional case, permitting direct comparison. Both structures were tested under identical solar irradiance conditions (1 sun, AM1.5G) for both front and back illumination, for three different strain-states (coherently strained, fully and partially relaxed). The non-polar case was also examined.

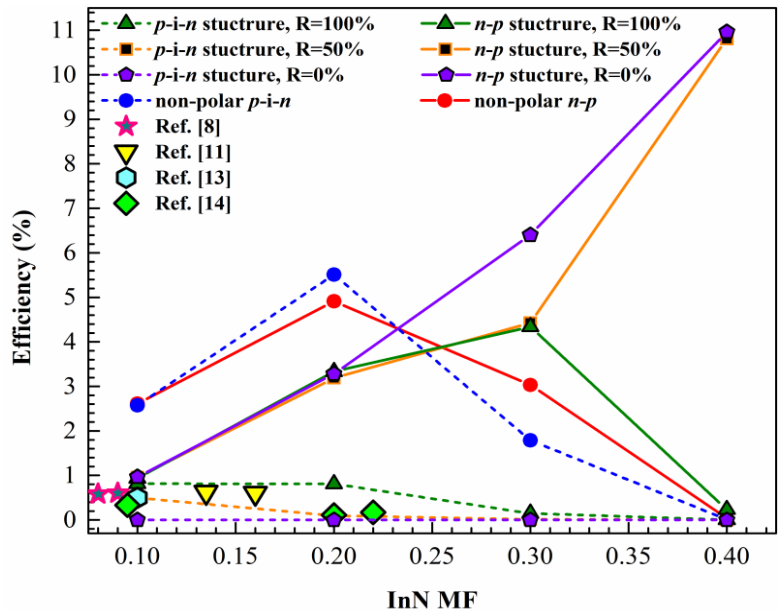

Fig.2. Calculated AM1.5G conversion efficiency versus alloy composition for $p$-i-$n$ and $n$-$p$ solar cells under different strain-states and when no polarization charges are taken into account. Typical reported measured efficiencies of $p$-GaN/i-InGaN/$n$-GaN are also plotted for comparison.

In Fig. 2 the conversion efficiency is presented as a function of InN MF for the two designs under various degrees of relaxation, as well as for the case where no polarization charges are considered. It should be noted that all values in Fig. 2 are for InGaN electron concentration $5 \times 10^{17}$ cm$^{-3}$. Also, the values for the polar structures are under the optimum illumination conditions, meaning that the *p*-i-*n* structure is more efficient when illuminated from the top, whereas the *n-p* when illuminated from the back. This can be explained by the measured wavelength-dependent refractive index curves illustrated in Fig. 1. To maximize the efficiency of a solar cell, the layer-stacking sequence should be from low to high refractive index values, for enhanced absorption and light trapping. In the *p*-i-*n* architecture the high refractive index material (i-InGaN) is between lower refractive index materials (*p* and *n*-GaN), so the direction of incident light is not so crucial. However, in the *p*-side down design under back illumination, the refractive index gradient (from low to high values) is beneficial for efficiency, whereas under front illumination the reflection losses are greater.

In non- polar structures, as depicted in Fig. 2, an increase in conversion efficiency with indium content was observed due to the bandgap lowering, which leads to enhanced absorption of the InGaN film. For InN MF greater than 0.2, the efficiency decreases abruptly. In such polarization-free structures, the efficiency drop with indium composition can be explained by the potential barrier at the GaN/InGaN hetero-interfaces, arisen from the large electron affinity difference between GaN and InN, which increases with indium content and thus impedes carrier transport across the device. In a polar *p*-i-*n* solar cell, polarization charges along with the inherent affinity-driven barrier at the GaN/InGaN interface deteriorate the photo-generated carrier collection, an effect which is more pronounced for higher InN MF (required for high efficiency III-nitride solar cells) or/and strain degrees [7]–[9]. In comparison to these theoretical predictions, in Fig.2 typical reported measured conversion efficiencies [8], [11], [13], [14] of *p*-GaN/i-InGaN/*n*-GaN devices, with relevant heterostructure characteristics, are shown. The good agreement observed emphasizes the fact that polarization-induced electric field is the major limiting factor in the current status of InGaN-based photovoltaic devices, rather than the material quality, as often attributed to.

Typical conversion efficiency values are shown in Fig.2, for partially and fully relaxed as well as coherently strained InGaN films. As seen, the indium content and the strain-state have the opposite effect on the efficiency when the *n*-on-*p* architecture is employed, reaching ~11% for the $In_{0.4}Ga_{0.6}N$ film under partial relaxation and coherently strained case. To elaborate on this huge efficiency difference between *p*-i-*n* and *n*-*p* structures, the energy band diagrams along with the three types of recombination considered (Shockley-Read-Hall (SRH), radiative and Auger) are plotted in Fig. 3, for partially relaxed InGaN layers with 40% indium content. As seen in Fig. 3(a), in the *p*-GaN/i-InGaN/*n*-GaN stacking sequence, the conduction and valence band in the absorbing layer are tilted in a direction detrimental for efficient carrier collection. Photo-generated electrons and holes drift towards the *p*- and *n*-region, respectively, and consequently recombine before contributing to the current. The origin of such band-bending is the polarization-induced electric filed, which not only compensates the normal built-in field of the junction since they have opposite directions, but severely overcomes it with its great magnitude upon increasing InN MF or/and decreasing plastic relaxation percentage of the films [9], [20], [21]. Also, in such designs, as the indium content or/and strain degree rises, the escalating positive polarization charges at the *p*-GaN/InGaN hetero-interface create an additional potential barrier. The conduction band at the hetero-interface is pulled down approaching the Fermi level and at high polarization charge densities eventually touches or ever lies below Fermi level, totally preventing carrier extraction [9]. A characteristic example describing the above problems and explaining the low

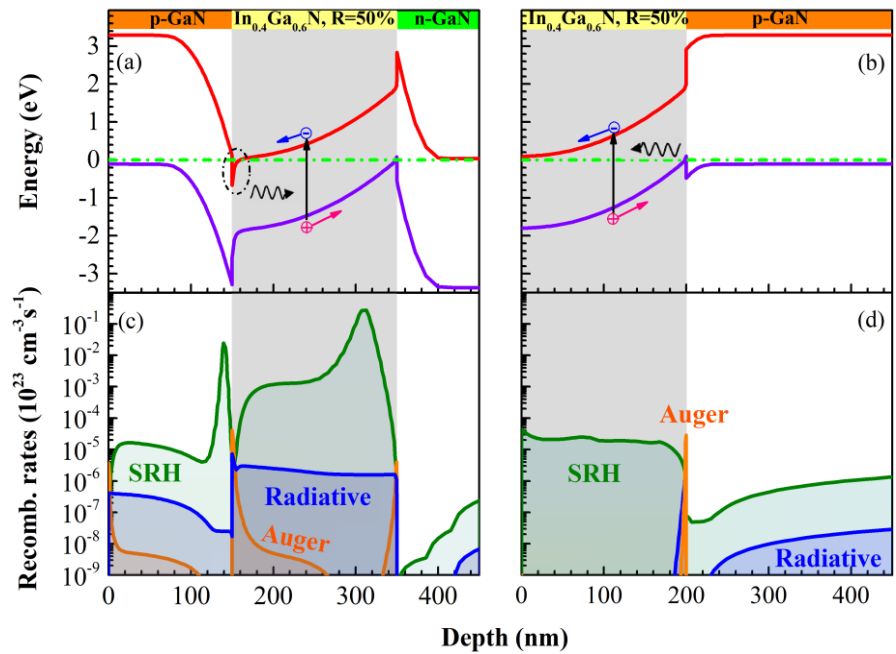

Fig.3. (a) Energy band diagram of a conventional *p*-i-*n* InGaN-based solar cell compared with (b) that of *n*-*p* structure. Their corresponding recombination rates are presented at (c) and (d), respectively.

performance of GaN/InGaN double heterojunction solar sell is presented in Fig. 3 (a) and (c).

In contrast with *p*-i-*n* double heterojunction, *n-p* design possesses completely different characteristics. Now, the direction of polarization-induced electric field is in the opposite direction compared to the typical stacking sequence, enhancing the normal build-in field of the junction as the InN MF or/and strain degree increases, and, thus, the energy band-bending is in favor of carrier collection, as illustrated in Fig. 3(b). Additionally, the negative polarization sheet charges at the *p*-GaN/InGaN hetero-interface, whose density is augmented with indium content and strain-state, diminish the potential barrier due to affinity difference, improving further the photo-generated carrier extraction. In Fig. 3(d), recombination rates are presented for the $In_{0.4}Ga_{0.6}N$ film case with R=50%. As demonstrated, the prevailing recombination mechanism, SRH, has a recombination rate at least three orders of magnitude lower than in the conventional structure, clarifying the enormous difference at the conversion efficiency values. However, a sharp drop in the efficiency is observed in Fig. 2, for 40% indium content film when it is fully relaxed. At such InN MF, the affinity-induced barrier is high enough and the polarization charge density at the interface is not adequate to reduce it, limiting the efficient carrier collection, as will be further investigated in the following section.

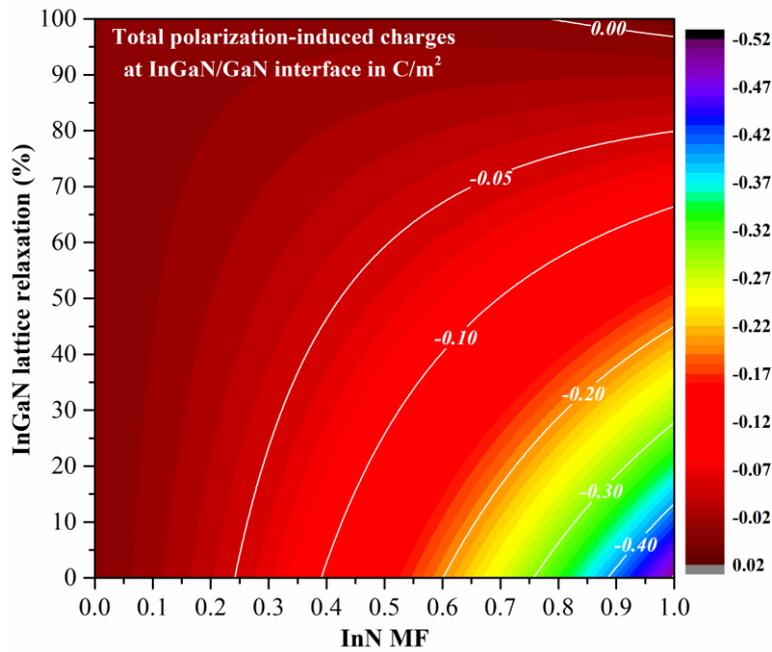

Fig.4. The calculated total polarization-induced charges at the InGaN/GaN interface versus InN MF and relaxation status based on [27].

To elaborate on the strength of the polarization-induced electric field in the InGaN layer, the total polarization bound charge density at the InGaN/GaN interface due to the difference is spontaneous polarizations and the InGaN piezoelectric contribution, is plotted in

Fig. 4, as a function of the alloy layer InN MF and lattice relaxation. As observed, the bound charges are negative in almost all cases. Therefore, for *n-p* architecture the induced field is aligned to the built-in field of the junction and consequently benefits the carrier collection efficiency. However, this effect is decreasing monotonically with lattice relaxation. The importance of R on the performance of high indium content solar cell modules will be examined in details in the next sections.

### B. Doping and Thickness of InGaN layer

Based on the previous results, high efficiency InGaN photovoltaics can be achieved by adopting a simple *n-p* architecture, avoiding complex structures and high *p*-type doping concentrations which are difficult to realize. In order to explore the true potential of this structure, additional simulations series were performed. The InN MF range was expanded up to 0.62 and the fully coherent InGaN films case was also considered. Different InGaN doping levels are studied, as well as different thicknesses, to reveal the optimum features that a high efficiency *n*-InGaN/*p*-GaN photovoltaic module should have.

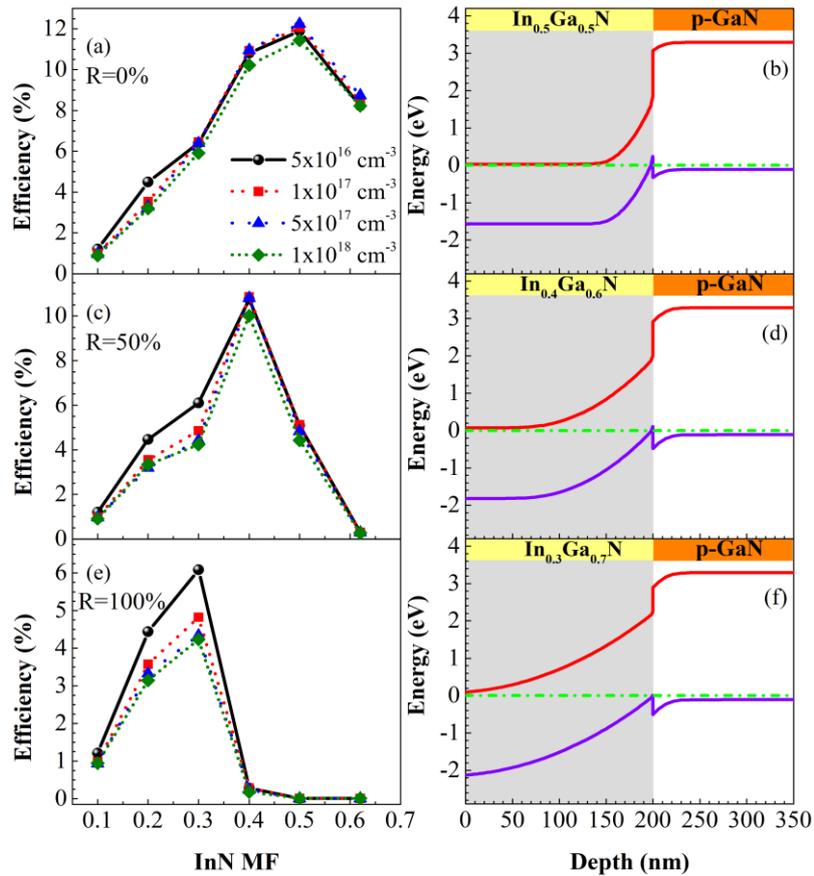

Fig.5. (a), (c), (e) Conversion efficiency values of the single heterojunction *n-p* InGaN solar cell versus InN MF at different InGaN electron concentrations for 200 nm thick film coherently strained, partially and fully relaxed, respectively. (b), (d), (f) Energy band diagrams corresponding to the highest efficiency recorded for each case presented at (a), (c) and (e).

The evolution of conversion efficiency for various InGaN doping concentration values and relaxation statuses upon increasing indium content of the films are illustrated in Fig. 5(a), (c) and (e) along with their corresponding band diagrams for the highest value for each case of (b), (d) and (f) respectively. As depicted in Fig. 5(a), (c) and (e) electron concentration of the InGaN absorbing layer does not play a significant role in the performance of the solar cell for the values studied here. This is another advantage of the *n-p* design. These structures are not relying on the doping dependent diffusion but on the field dependent drift process. Since the total electric field of the junction takes huge values mainly due to the polarization, doping does not seriously affect conversion efficiency. By adopting this design, the requirement for low *n*-doping levels, which are challenging in InGaN epitaxial layers, is relaxed.

Relaxation status is one of the most crucial parameters that dictate *n-p* solar cell performance. When films are fully strained, the efficiency monotonically increases with InN MF and reaches approximately 12% for 50% indium composition. For higher indium contents, the magnitude of valence band offset is sufficient enough to prevent carrier collection resulting to efficiency deterioration. As the InGaN films relax the effect of the affinity-driven potential barrier is more prominent, resulting to efficiency degradation and to a shift of the maximum values towards lower indium compositions. On partially and fully relaxed cases, the polarization charges at the hetero-interface have the adequate density to moderate the valence band offset up to 0.4 and 0.3 InN MF values respectively. Beyond that indium contents affinity barrier prevails, hindering the photo-generated carrier collection.

Another parameter of utmost importance for the efficiency of a solar cell is the thickness of the absorbing layer. In Fig. 6, the maximum conversion efficiency values recorded for each case of indium content are presented as a function of InGaN thickness. The doping concentration of the films in this case was $5 \times 10^{16}$ cm$^{-3}$ for the non-polar structures whereas for the polar it was $1 \times 10^{17}$ cm$^{-3}$. As illustrated, the conversion efficiency in all cases increases upon elevated absorbing layer thickness. For indium compositions 10 and 20%, non-polar *n*-InGaN/*p*-GaN solar cells are performing better compared to their polar counterparts but, in order to absorb a wider part of the solar spectrum, films with higher InN MF are required. In such cases, the conversion efficiency can be enhanced only by using coherently strained layers. Under this relaxation status, efficiencies are greater than 8%, for In-rich films of 200 nm or more. Especially for a coherently strained 300 nm thick In$_{0.5}$Ga$_{0.5}$N film, efficiency reaches 14%.

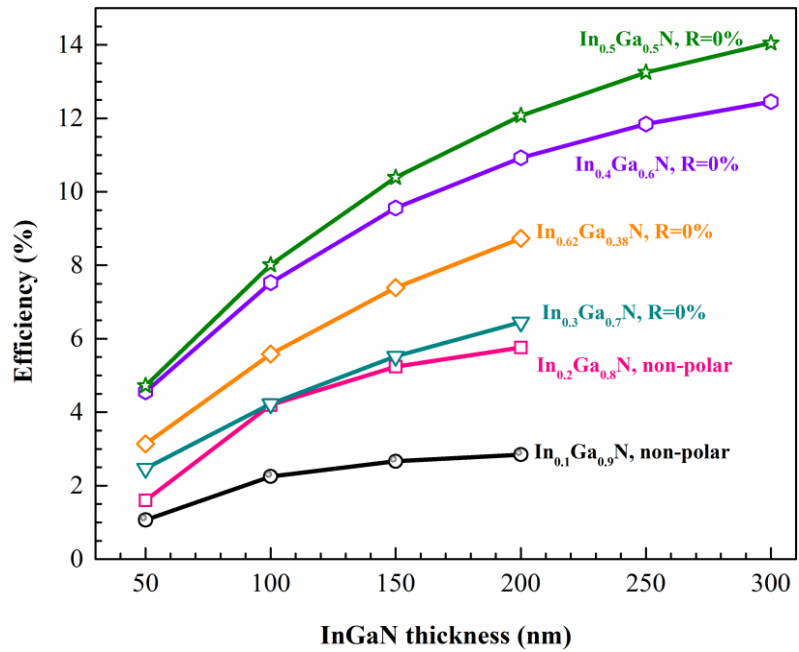

Fig.6. Best conversion efficiency values achieved by employing *n-p* architecture versus InGaN film thickness for different InGaN alloy compositions.

## 4. CONCLUSIONS

The photovoltaic operation of single (*n*-InGaN/*p*-GaN) and double (*p*-GaN/*i*-InGaN/*n*-GaN) heterojunction structures was theoretically studied. The effects of InN mole fraction, layer thickness, strain state and doping levels on the conversion efficiency of the cells were considered. The importance of the polarization charges in reducing the barrier at the InGaN/GaN interface and in increasing the photo-generated carrier collection is revealed. In the case of single heterojunctions, the efficiency monotonically increases for indium mole fractions up to 0.5 when the InGaN film is coherently strained on the GaN substrate, whereas it could be maintained at high values even if the InGaN layer is partially relaxed for mole fractions up to 0.4. Also, it was found that the InGaN n-doping concentration does not significantly affect the photovoltaic efficiency, whereas a conversion efficiency up to 14% can be reached for the case of a 300 nm, coherently strained $In_{0.5}Ga_{0.5}N$ absorbing layer.

**ACKNOWLEDGMENTS**

This work co-financed by the European Union (European Social Fund–ESF) and Greek national funds through the Operational Program "Education and Lifelong Learning" of the National Strategic Reference Framework (NSRF)-Research Funding Program: THALES, project "NitPhoto".

References

[1] J. Wu, W. Walukiewicz, W. Shan, K. M. Yu, J. W. Ager, S. X. Li, E.E. Haller, H. Lu, and W. J. Schaff, "Temperature dependence of the fundamental band gap of InN," *J. Appl. Phys.*, vol. 94, no. 7, pp. 4457–4460, Oct. 2003.

[2] T. Kawashima, H. Yoshikawa, S. Adachi, S. Fuke, and K. Ohtsuka, "Optical properties of hexagonal GaN," *J. Appl. Phys.*, vol. 82, no. 7, pp. 3528–3535, Oct. 1997.

[3] R. Singh, D. Doppalapudi, T. Moustakas, L. Romano, "Phase separation in InGaN thick films and formation of InGaN/GaN double heterostructures in the entire alloy composition," *Appl. Phys. Lett.*, vol. 70, no. 9, pp. 1089–1091, Mar. 1997.

[4] A. David, M. Grundmann, "Influence of polarization fields on carrier lifetime and recombination rates in InGaN-based light-emitting diodes," *Appl. Phys. Lett.*, vol. 97, pp. 033501-1–033501-3, May 2010.

[5] J. Wu, W. Walukiewicz, K. M. Yu, W. Shan, and J. W. Ager III, "Superior radiation resistance of In$_{1-x}$Ga$_x$N alloys: Full solar-spectrum photovoltaic material system," *J. Appl. Phys.*, vol. 94, no. 10, pp. 6477–6482, Jun. 2013.

[6] G. F. Brown, J. W. Ager, III, W. Walukiewicz, and J. Wu, "Finite element simulations of compositionally graded InGaN solar cells," *Solar Energy Mater. Solar Cells*, vol. 94, pp. 478–483, Mar. 2010.

[7] Z. Q. Li, M. Lestradet, Y. G. Xiao, S. Li, "Effects of polarization charge on the photovoltaic properties of InGaN solar cells," *Phys. Status Solidi (a)*, vol. 208, pp. 928-931, no. 4, Dec. 2010.

[8] J. J. Wierer, Jr., A. J. Fischer, and D. D. Koleske, "The impact of piezoelectric polarization and nonradiative recombination on the performance of (0001) face GaN/InGaN photovoltaic devices," *Appl. Phys. Lett.*, vol. 96, no. 5, pp. 051107-1–051107-3, Feb. 2010.

[9] J.-Y. Chang and Y.-K. Kuo, "Numerical study on the influence of piezoelectric polarization on the performance of p-on-n (0001)-face GaN/InGaN p-i-n solar cells," *IEEE Electron Device Lett.*, vol. 32, no.7, pp. 937–939, Jul. 2011.

[10] C. A. M. Fabien *et al.*, "Simulations, practical limitations and novel growth technology for InGaN-based solar cells," *IEEE J. Photovolt.*, vol. 4, no. 2, pp. 601–606, Mar. 2013.

[11] X. Cai *et al.*, "Investigation of InGaN p-i-n homojunction and heterojunction solar cells," *IEEE Photon. Technol. Lett.*, vol. 25, no. 1, pp. 59–62, Jan. 2013.

[12]   S. Valdueza-Felip *et al.*, "P-i-n InGaN homojunctions (10–40% In) synthesized by plasma-assisted molecular beam epitaxy with extended photoresponse to 600 nm," *Solar Energy Mater. Solar Cells*, vol. 160, pp. 355–360, Feb. 2017.

[13]   X. Zheng *et al.*, "High-quality InGaN/GaN heterojunctions and their photovoltaic effects", *Appl. Phys. Lett.*, vol. 93, pp. 261108-1–261108-3, Dec. 2008.

[14]   C. A. M. Fabien, A. Maros, C. B. Honsberg, W. A. Doolitle, "III-nitride double heterojunction solar cells with high In-content InGaN absorbing layers: comparison of large-area and small-area devices," *IEEE J. Photovolt.*, vol. 6, pp. 460 - 464, Dec. 2015.

[15]   O. Jani, I. Ferguson, C. Honsberg, and S. Kurtz, "Design and characterization of GaN/InGaN solar cells," *Appl. Phys. Lett.*, vol. 91, no. 13, pp. 132117-1–132117-3, Sep. 2007.

[16]   C. J. Neufeld, N. G. Toledo, S. C. Cruz, M. Iza, S. P. DenBaars, and U. K. Mishra, "High quantum efficiency InGaN/GaN solar cells with 2.95 eV band gap", *Appl. Phys. Lett.*, vol. 93, pp. 143502-1–143502-3, Oct. 2008.

[17]   B. W. Liou, "Design and fabrication of $In_xGa_{1-x}N$/GaN solar cells with a multiple-quantum-well structure on SiCN/Si (111) substrates", *Thin Solid Films*, vol. 530, pp. 1084–1090, Nov. 2011

[18]   N. G. Young *el al.*, "High-performance broadband optical coatings on InGaN/GaN solar cells for multijunction device integration," *Appl. Phys. Lett.*, vol. 104, pp. 163902-1–163902-4, Apr. 2014.

[19]   C. J. Neufeld *et al.*, "Effect of doping and polarization on carrier collection in InGaN quantum well solar cells," *Appl. Phys. Lett.*, vol. 98, pp. 243507-1– 243507-3, Jun. 2011.

[20]   Y.-K. Kuo, J.-Y. Chang, and Y.-H. Shih, "Numerical study of the effects of hetero-interfaces, polarization charges, and step-graded interlayers on the photovoltaic properties of (0001) face GaN/InGaN p-i-n solar cell," *IEEE J. Quantum Electron.*, vol. 48, no. 3, pp. 367–374, Mar. 2012.

[21]   J.-Y. Chang, S.-H. Yen, Y.-A. Chang, and Y.-K. Kuo, "Simulation of high-efficiency GaN/InGaN p-i-n solar cell with suppressed polarization and barrier effects," *IEEE J. Quantum Electron.*, vol. 49, no. 1, pp. 17–23, Jun. 2013.

[22]   J. R. Dickerson, K. Pantzas, A. Ougazzaden, and P. L. Voss, "Polarization-induced electric fields make robust n-GaN/i-InGaN/p-GaN solar cells", *IEEE Electron. Device Lett.*, Vol. 34, no. 3, Mar. 2013

[23]   M. L. Reed *et al.*, "n-InGaN/p-GaN single heterostructure light emitting diode with p-side down", *Appl. Phys. Lett.*, vol. 93, pp. 133505-1– 133505-3, Sep. 2008.

[24] APSYS by Crosslight Software Inc., Burnaby, BC, Canada.

[25] S. L. Chuang and C. S. Chang, "k·p method for strained wurtzite semiconductors," *Phys. Rev. B*, vol. 54, no. 4, pp. 2491–2504, Jul. 1996.

[26] S. L. Chuang and C. S. Chang, "A band-structure model of strained quantum-well wurtzite semiconductors," *Semicond. Sci. Technol.*, vol. 12, no. 3, pp. 252–263, Mar. 1997.

[27] V. Fiorentini, F. Bernardini, and O. Ambacher, "Evidence for nonlinear macroscopic polarization in III–V nitride alloy heterostructures," *Appl. Phys. Lett.*, vol. 80, pp. 1204–1206, Feb. 2002.

[28] J. Simon, V. Protasenko, C. Lian, H. Xing, and D. Jena, "Polarization-induced hole doping in wide band-gap uniaxial semiconductor heterostructures," *Science*, vol. 327, pp. 60–64, Jan. 2010.

[29] Z. Z. Bandic, P. M. Bridger, E. C. Piquette, and T. C. McGill, "Minority carrier diffusion length and lifetime in GaN," Appl. Phys. Lett., vol. 72, pp. 3166–3168, Jun. 1998.

[30] F. Chen, A. N. Cartwright, H. Lu, and W. J. Schaff, "Temperature dependence of carrier lifetimes in InN," Phys. Status Solidi A, vol. 202, pp. 768–772, Apr. 2005.

[31] J. F. Muth *et al.*, "Absorption coefficient, energy gap, exciton binding energy, and recombination lifetime of GaN obtained from transmission measurements," *Appl. Phys. Lett.*, vol. 71, no. 18, pp. 2572–2574, Jun. 1998.

[32] Chen, A. N. Cartwright, H. Lu, and W. J. Schaff, "Time-resolved spectroscopy of recombination and relaxation dynamics in InN," *Appl. Phys. Lett.*, vol. 83, no. 24, pp. 4984–4986, Dec. 2003.

[33] K. T. Delaney, P. Rinke, and C. G. Van de Walle, "Auger recombination rates in nitrides from first principles," *Appl. Phys. Lett.*, vol. 94, no. 19, pp. 191109-1–191109-3, Dec. 2003.

[34] J. Piprek, *Semiconductor Optoelectronic Devices: Introduction to Physics and Simulation*. San Diego, CA: Academic, 2003, pp. 61–66.

[35] E. Papadomanolaki, C. Bazioti, S. A. Kazazis, M. Androulidaki, G. P. Dimitrakopulos, E. Iliopoulos, "Molecular beam epitaxy of thick InGaN(0001) films: Effects of substrate temperature on structural and electronic properties", *J. Cryst. Growth*, vol. 437, pp. 20–25, Mar. 2016.

[36]  I. Gorczyca, S. P. Łepkowski, and T. Suski, "Influence of indium clustering on the band structure of semiconducting ternary and quaternary nitride alloys", *Phys. Rev. B*, vol.80, pp. 075202-1–075202-11, Aug. 2009.

[37]  S.-H. Wei and A. Zunger, "Valence band splittings and band offsets of AlN, GaN, and InN", *Appl. Phys. Lett.*, vol. 69, no. 18, pp. 2719–2721, Oct. 1996.

[38]  B. N. Pantha, H. Wang, N. Khan, J. Y. Lin, and H. X. Jiang, "Origin of background electron concentration in In$_x$Ga$_{1-x}$N alloys", *Phys. Rev. B*, vol. 84, pp. 075327-1–075327-6, Aug. 2011.

[39]  I. Ho and G. B. Stringfellow, "Solid phase immiscibility in GaInN", *Appl. Phys. Lett.*, vol. 69, no. 18, pp.2701-2703, Sep. 1996.

[40]  D. Holec, P. M. F. J. Costa, M. J. Kappers, C. J. Humphreys, "Critical thickness calculations for InGaN/GaN", *J. Cryst. Growth*, vol. 303, pp.314-317, May 2007.

[41]  S. Pereira *et al.*, "Structural and optical properties of InGaN/GaN layers close to the critical layer thickness", *Appl. Phys. Lett.*, vol. 81, no.7, pp.1207-1209, Aug. 2002.

[42]  W. Zhao, L. Wang, J. Wang, Z. Hao, Y. Luo, "Theoretical study on critical thicknesses of InGaN grown on (0 0 0 1) GaN", *J. Cryst. Growth*, vol. 327, pp.202-204, Jul. 2011.